\documentclass[cits]{PoS}

\usepackage[utf8]{inputenc}
\usepackage{amsmath,amssymb,fontenc,enumerate,times,mathptmx,graphicx,float}

\usepackage[nodisplayskipstretch]{setspace}

\newcommand{\RBRC}{
  RIKEN BNL Research Center,
  Brookhaven National Laboratory,
  Upton, New York 11973,
  USA}

\newcommand{\UCONN}{
  Physics Department,
  University of Connecticut,
  Storrs, Connecticut 06269-3046,
  USA}

\newcommand{\NAGOYA}{
  Department of Physics,
  Nagoya University,
  Nagoya 464-8602,
  Japan}
  
\newcommand{\NISHINA}{
  Nishina Center,
  RIKEN,
  Wako, Saitama 351-0198,
  Japan}

\newcommand{\BNL}{
  Physics Department,
  Brookhaven National Laboratory,
  Upton, New York 11973,
  USA}

\newcommand{\CU}{
  Physics Department,
  Columbia University,
  New York, New York 10027,
  USA}

\title{Hadronic Light by Light Contributions to the Muon Anomalous Magnetic Moment With Physical Pions}

\ShortTitle{Light-by-Light of Muon $g-2$}

\author{\speaker{Luchang Jin}\\
        \CU\\
        E-mail: \email{lj2289@columbia.edu}}

\author{Thomas Blum\\
{\UCONN}\\
{\RBRC}}

\author{Norman Christ\\
  {\CU}}

\author{Masashi Hayakawa\\
{\NAGOYA}\\
{\NISHINA}}

\author{Taku Izubuchi\\
{\BNL}\\
{\RBRC}}

\author{Christoph Lehner\\
{\BNL}}

\abstract{
The current measurement of muonic $g - 2$ disagrees with the theoretical
calculation by about 3 standard deviations. Hadronic vacuum polarization (HVP)
and hadronic light by light (HLbL) are the two types of processes that
contribute most to the theoretical uncertainty. The current value for HLbL is
still given by models. I will describe results from a first-principles lattice
calculation with a 139 MeV pion in a box of 5.5 fm extent. Our current
numerical strategies, including noise reduction techniques, evaluating the
HLbL amplitude at zero external momentum transfer, and important remaining
challenges, in particular those associated with finite volume effects, will be
discussed. }

\FullConference{The 33rd International Symposium on Lattice Field Theory\\
	14 -18 July 2015\\
	Kobe International Conference Center, Kobe, Japan}

\begin{document}

\setlength{\abovedisplayskip}{0.2cm}
\setlength{\belowdisplayskip}{0.2cm}

\section{Introduction}

\begin{figure}[H]
  \begin{center}
    \resizebox{0.3\columnwidth}{!}{\includegraphics{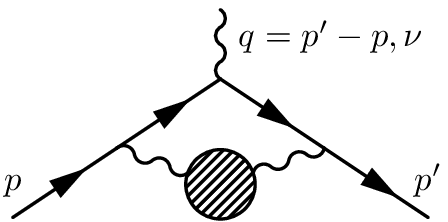}}
    \resizebox{0.3\columnwidth}{!}{\includegraphics{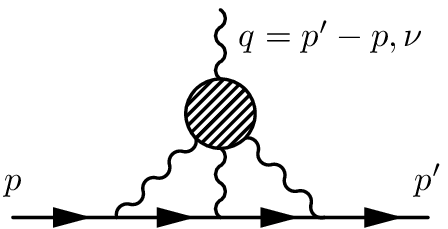}}
  \end{center}
  
\vspace{-0.6cm}
  \caption{\label{fig:hvp-hlbl}(Left) Hadronic vacuum polarization (HVP).
  (Right) Hadronic light-by-light (HLbL).}
\end{figure}

The anomalous magnetic moment of muon, $a_{{\mu}}$, can be defined in
terms of the photon-muon vertex function:\vspace{-0.1cm}
\begin{eqnarray}
  \langle \mathbf{p}', s' | j_{\nu}
  (\mathbf{x}_{\ensuremath{\operatorname{op}}} =\ensuremath{\boldsymbol{0}}) |
  \mathbf{p}, s \rangle & = & - e \bar{u}_{s'} (\mathbf{p}') \left[ F_1 (q^2)
  \gamma_{\nu} + i \frac{F_2 (q^2)}{4 m} [\gamma_{\nu}, \gamma_{\rho}]
  q_{\rho} \right] u_s (\mathbf{p}),  \label{eq:vertex-func}
\end{eqnarray}
where $F_1 (q^2 = 0) = 1$, $F_2 (q^2 = 0) = (g_{{\mu}} - 2) / 2 \equiv
a_{{\mu}}$, $q = p' - p$. The formula is written in a Euclidean gamma
matrix convention, $[\gamma_{{\mu}}, \gamma_{\nu}] = \delta_{{\mu},
\nu}$. Its value has been measured very precisely by BNL E821
{\cite{Bennett:2006fi}}. It can also be calculated theoretically to great
precision {\cite{Blum:2013xva}}. The three standard deviation $(287 \pm 80)
\times 10^{- 11}$ difference between experiment and theory makes muon $g - 2$
a very interesting quantity. Much more accurate experiments, Fermilab E989
{\cite{Gray:2015qna}} and J-PARC E34 {\cite{Nishimura:2015pra}}, are expected
to start in a few years, so a more accurate theoretical determination will be
necessary. Figure \ref{fig:hvp-hlbl} shows the two diagrams that are the major
sources of the theoretical uncertainty.\vspace{-0.3cm}

\vspace{-0.4cm}

In this paper, we will only discuss the lattice calculation of connected
hadronic light-by-light amplitude. Previously this quantity has only been
calculated using models {\cite{Prades:2009tw}}. Attempts using lattice QCD
were begun by T. Blum, M. Hayakawa, and T. Izubuchi more than 5 years ago
{\cite{Hayakawa:2005eq,Blum:2014oka}}. We have improved the
methodology dramatically, as described in Ref
{\cite{Jin:2015eua,Blum:2015gfa}}, which leads to a reduction in
statistical errors by more than an order of magnitude. Since much of the
material that was presented at LATTICE 2015 has now appeared in a long paper
{\cite{Blum:2015gfa}}, this proceedings is devoted to an expanded discussion
of three topics that were only briefly presented during the conference: a) The
sampling strategy that concentrated the result on the more easily evaluated,
short distance region in position space. b) A new proposal to perform the QCD
and QED portions of the HLbL calculation in different space-time volumes and
c) First numerical results from a physical-quark-mass, $48^3 \times 96$ study.

\section{Sampling Strategy}

\begin{figure}[H]
  \begin{center}
    \resizebox{0.25\columnwidth}{!}{\includegraphics{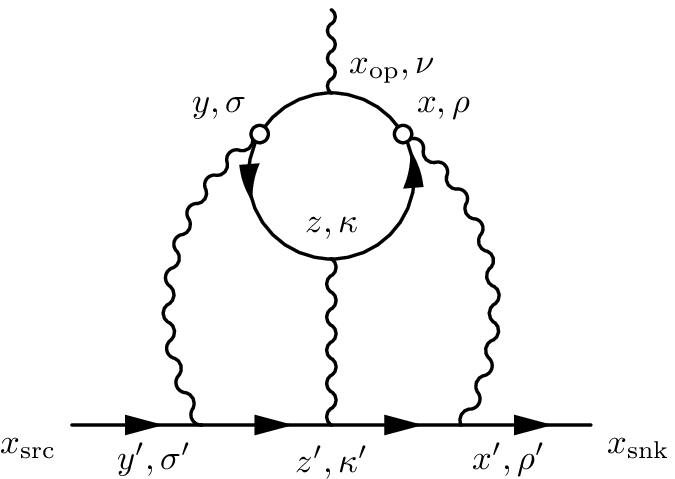}}
    \resizebox{0.25\columnwidth}{!}{\includegraphics{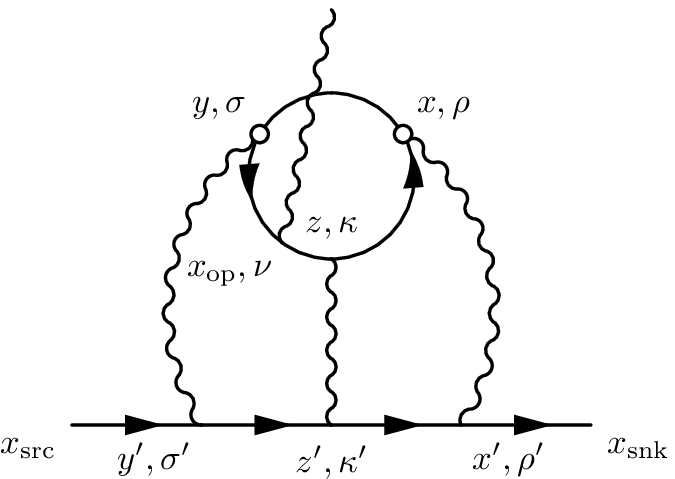}}
    \resizebox{0.25\columnwidth}{!}{\includegraphics{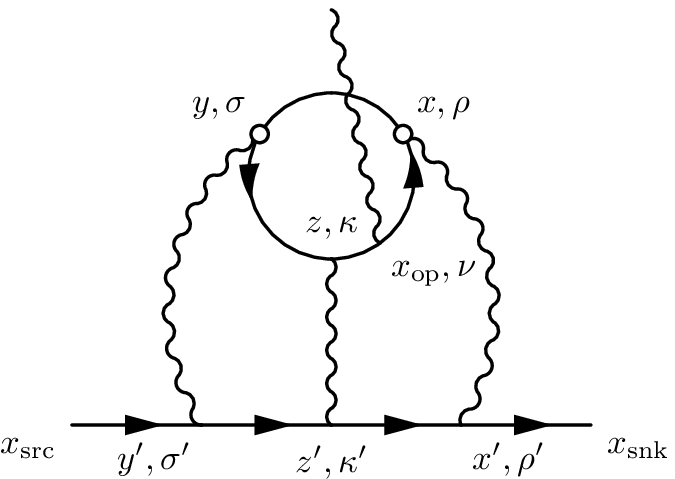}}
  \end{center}
  
\vspace{-0.7cm}
  \caption{\label{fig:c-3}Diagrams showing the three different ways of
  inserting the external photon when the vertices $x$ and $y$ are fixed. For
  each of these three diagrams there are five other possible permutations of
  the connections between the three internal photons and the muon line that
  are not shown.}
\end{figure}

We start the discussion by repeating our final formula for evaluating the
light-by-light contribution to $F_2 (q^2 = 0)$ obtained in Refs
{\cite{Jin:2015eua,Blum:2015gfa}}.
\begin{eqnarray}
  \frac{F_2 (q^2 = 0)}{m}  \frac{(\sigma_{s' s})_i}{2} & = & \sum_{r,
  \tilde{z}} \mathfrak{Z} \left( \frac{r}{2}, - \frac{r}{2}, \tilde{z} \right)
  \sum_{\tilde{x}_{\text{op}}}  \frac{1}{2} \epsilon_{i j k} \left(
  \tilde{x}_{\text{op}} \right)_j \cdot i \bar{u}_{s'} (\vec{0})
  \mathcal{F}^C_k \left( \frac{r}{2}, - \frac{r}{2}, \tilde{z},
  \tilde{x}_{\text{op}} \right) u_s (\vec{0}) . 
  \label{eq:f2-lbl-moment-short-z}
\end{eqnarray}
where $(\sigma_{s' s})_i = \bar{u}_{s'} (\vec{0}) \Sigma_i u_s (\vec{0})$ are
the conventional Pauli matrices, $\Sigma_k = \frac{1}{4 i} \epsilon_{i j k}
[\gamma_i, \gamma_j]$ and the weight function ``$\mathfrak{Z}$'' is defined
below but could be replaced by $1$. The integration variables are related to
the coordinates in Figure \ref{fig:c-3} by the following equations: $r = x -
y$, $\tilde{z} = z - (x + y) / 2$, $\tilde{x}_{\text{op}} = x_{\text{op}} - (x
+ y) / 2$, while
\begin{eqnarray}
  \mathcal{F}^C_{\nu} \left( x, y, z, x_{\text{op}} \right) & = & \frac{1}{3}
  \mathcal{F}_{\nu} \left( x, y, z, x_{\text{op}} \right) + \frac{1}{3}
  \mathcal{F}_{\nu} \left( y, z, x, x_{\text{op}} \right) + \frac{1}{3}
  \mathcal{F}_{\nu} \left( z, x, y, x_{\text{op}} \right), 
  \label{eq:lbl-amp-c}
\end{eqnarray}
\vspace{-0.6cm}

\vspace{-0.4cm}

The amplitude $\mathcal{F}_{\nu} \left( x, y, z, x_{\text{op}} \right)$ is
given by:
\begin{eqnarray}
  &  & \mathcal{F}_{\nu} \left( x, y, z, x_{\text{op}} \right) 
  \label{eq:lbl-amp}\\
  & = & \left\langle - \sum_{q = u, d, s} (e_q / e)^4
  \ensuremath{\operatorname{tr}} \left[ \gamma_{\rho} S_q (x, z)
  \gamma_{\kappa} S_q (z, y) \gamma_{\sigma} S_q \left( y, x_{\text{op}}
  \right) \gamma_{\nu} S_q \left( x_{\text{op}}, x \right) \right]
  \right\rangle_{\text{QCD}} \nonumber\\
  & \cdot & \sum_{x', y', z'} (- i e)^6 G_{\rho \rho'} (x, x') G_{\sigma
  \sigma'} (y, y') G_{\kappa \kappa'} (z, z') \nonumber\\
  & \cdot & e^{m_{{\mu}}  \left( t_{\text{snk}} - t_{\text{src}} \right)}
  \sum_{\vec{x}_{\text{snk}}, \vec{x}_{\text{src}}} \left[ S_{{\mu}}
  \left( x_{\text{snk}}, x' \right) \gamma_{\rho'} S_{{\mu}} (x', z')
  \gamma_{\kappa'} S_{{\mu}} (z', y') \gamma_{\sigma'} S_{{\mu}}
  \left( y', x_{\text{src}} \right) \right. \nonumber\\
  & \cdot & \left. + S_{{\mu}} \left( x_{\text{snk}}, z' \right)
  \gamma_{\kappa'} S_{{\mu}} (z', x') \gamma_{\rho'} S_{{\mu}} (x',
  y') \gamma_{\sigma'} S_{{\mu}} \left( y', x_{\text{src}} \right) +
  \text{other 4 permutations} \right], \nonumber
\end{eqnarray}
where $e_u / e = 2 / 3$, $e_d / e = e_s / e = - 1 / 3$.

We perform the sum over $r = x - y$ by sampling a few values on each QCD
configuration and use point source propagators at $x$ and $y$. The sum over
$\tilde{x}_{\text{op}}$ can be performed as a sequential source, $\tilde{z}$
is summed over as a sink. Since $x$ and $y$ are connected by quark lines, it
can be expected that the major contribution to $g - 2$ comes from the region
where $r$ is small. We accommodate this by using importance sampling, that is
we sample the small $r$ region more often, then divide the sample result by
the probability. We even perform a complete sum, up to discrete symmetries, in
the region where $r \leqslant r_{\mathrm{\max}}$. $r_{\text{max}}$ is usually
chosen to be $5$ in lattice units in our numerical simulations.

In our primary calculation we choose $\mathfrak{Z} (x, y, z)$ to be given by
\begin{eqnarray}
  \mathfrak{Z} (x, y, z) & = & \left\{ \begin{array}{ll}
    3 & \text{if } | x - y | < | x - z | \text{ and } | x - y | < | y - z |\\
    3 / 2 & \text{if } | x - y | = | x - z | < | y - z | \text{ or } | x - y |
    = | y - z | < | x - z |\\
    1 & \text{if } | x - y | = | x - z | = | y - z |\\
    0 & \text{otherwise}
  \end{array} \right. .  \label{eq:3-mult}
\end{eqnarray}
This choice of $\mathfrak{Z}$ restricts the summation region for $z$, in such
a way as to suppress the contribution from the large $r$ region. This way, the
connected $\pi^0$ exchange contribution would be completely captured by the
small $r$ region if we sum up to the size of the pion, instead of the Compton
wave length of the pion. We may also try the opposite of this choice, which
can provide more information about QCD finite-volume effects, and the size of
the long-distance $\pi^0$-exchange contribtion to the light-by-light process:
\begin{eqnarray}
  \mathfrak{Z}' (x, y, z) & = & \left\{ \begin{array}{ll}
    3 & \text{if } | x - y | > | x - z | \text{ and } | x - y | > | y - z |\\
    3 / 2 & \text{if } | x - y | = | x - z | > | y - z | \text{ or } | x - y |
    = | y - z | > | x - z |\\
    1 & \text{if } | x - y | = | x - z | = | y - z |\\
    0 & \text{otherwise}
  \end{array} \right. .  \label{eq:3-mult-long-z}
\end{eqnarray}
With this choice, in the large $r$ region, most of the contribution should
come from the connected $\pi^0$ exchange. How rapidly this contribution decays
could give us a hint about the size of the QCD finite-volume effect, which
comes from the quark propagators evaluated within the finite size lattice.

\section{QCD Box Inside a Larger QED Box}

\begin{figure}[H]
  \begin{center}
    \resizebox{0.5\columnwidth}{!}{\includegraphics{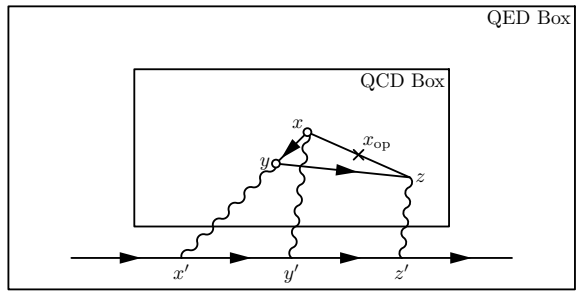}}
  \end{center}
  
\vspace{-0.6cm}
  \caption{\label{fig:box-in-box}QCD box inside QED box illustration.}
\end{figure}

The QCD finite volume effects are exponentially suppressed in the linear size
of the lattice volume times $m_{\pi}$, the energy of lowest energy eigen-state
of QCD. There are also QED finite volume effects, which are caused by the
photon and the muon propagators being evaluated within that finite volume, and
the summations over $\vec{x}_{\text{src}}$, $\vec{x}_{\text{snk}}$, $x'$,
$y'$, $z'$ being controlled to lie within that finite volume. Because the
photon is massless, the QED finite volume effects are power like, similar to
many other lattice calculations involving QED. In this particular case, these
QED finite volume effects scale like $1 / L^2$ as discussed in Ref
{\cite{Jin:2015eua,Blum:2015gfa}}. These QED finite volume effects can
be reduced by evaluating the photon and muon propagators and performing the
summation over $\vec{x}_{\text{src}}$, $\vec{x}_{\text{snk}}$, $x'$, $y'$,
$z'$ in Eq. (\ref{eq:lbl-amp}) in a larger volume compared to that in which
the QCD part of the calculation is performed. We refer to the former as the
QED box and the latter as the QCD box, see Figure \ref{fig:box-in-box}. With
our current light-by-light evaluation strategy, the computations for the quark
and the muon propagators are almost independent. We can compute the
light-by-light process for a few different QED boxes without recomputing the
quark part. One can then extrapolate to infinite volume based on these results
from different QED boxes. Since the quark part is the same, we expect there
will exist strong correlations between these results, which would benefit the
extrapolation. In principle, one could evaluate the muon and photon
propagators using the continuum formulae and perform the coordinate-space QED
summation in infinite volume directly, thus completely eliminating this
$\mathcal{O} (1 / L^2)$ finite volume effect. In fact, this is exactly the
strategy for the HVP calculation, where the usual approach
{\cite{Blum:2002ii}} can be viewed as substituting the finite-volume result
for $\Pi (q^2)$ into one- or two-loop QED calculations performed in infinite
volume. At this point, one can see that computing the QED part of the diagram
in a larger, possibly infinite, QED box is a quite general idea, and could be
applied in many (but not necessarily all) other lattice QCD calculations
involving QED. In some cases, like the QED mass-splittings, all one may need
to evaluate is the photon propagator in infinite volume. Christoph Lehner
talks about this in greater details in his talk at LATTICE 2015.
\vspace{-0.7cm}

\vspace{-0.4cm}

As discussed above, the finite volume effects of the light-by-light
calculation come from two sources. One source is the quark loop is evaluated
using a finite QCD box, which is exponentially suppressed in the size of the
QCD box. The other source is the photon and muon propagators evaluated in the
QED box, which scales like $1 / L^2$ as discussed in Ref
{\cite{Jin:2015eua,Blum:2015gfa}}. Here, we the strategy of making the
QED box larger than QCD box by a calculation in $16^3$ and $24^3$ lattice
volumes {\cite{Allton:2007hx,Allton:2008pn}} but with the same lattice
spacing and pion mass. All computations are performed on 14 configurations
separated by $200$ MD time unit and the results are shown in Table
\ref{fig:confs}.\vspace{-0.2cm}

\begin{table}[H]
  \begin{center}
    \begin{tabular}{ccccc}
      Ensemble & QCD Size & QED Size & $t_{\text{snk}} - t_{\text{src}}$ &
      $\frac{F_2 (q^2 = 0)}{ (\alpha / \pi)^3}$\\
      16I {\cite{Allton:2007hx}} & $16^3 \times 32$ & $16^3 \times 32$ & $16$
      & $0.1158 (8)$\\
      24I {\cite{Allton:2008pn}} & $24^3 \times 64$ & $24^3 \times 64$ & $32$
      & $0.2144 (27)$\\
      16I-24 {\cite{Allton:2007hx}} & $16^3 \times 32$ & $24^3 \times 64$ &
      $32$ & $0.1674 (22)$
    \end{tabular}
  \end{center}
  
\vspace{-0.6cm}
  \caption{\label{fig:confs}Finite volume effects studies. $a^{- 1} = 1.747
  \mathrm{\ensuremath{\operatorname{GeV}}}$, $m_{\pi} = 423
  \mathrm{\ensuremath{\operatorname{MeV}}}$, $m_{{\mu}} = 332
  \mathrm{\ensuremath{\operatorname{MeV}}}$.}
\end{table}

\vspace{-0.8cm}

\begin{figure}[H]
  \begin{center}
    \resizebox{0.45\columnwidth}{!}{\includegraphics{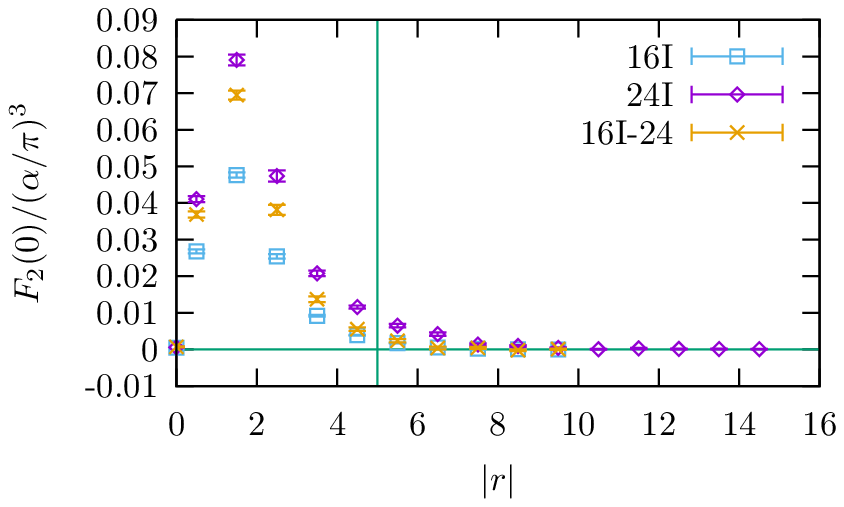}}
    \resizebox{0.45\columnwidth}{!}{\includegraphics{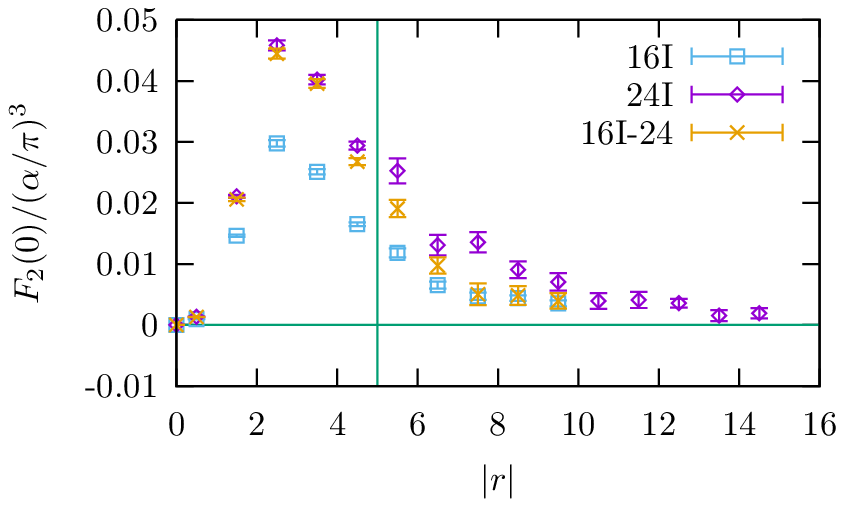}}
  \end{center}
  
\vspace{-0.8cm}
  \caption{\label{fig:bin-distance-16I-24I}Above plots show histograms of the
  contribution to $F_2$ from different separations $| r | = | x - y |$. The
  sum of all these points gives the final result for $F_2$. The vertical lines
  at $| r | = 5$ in the plots indicate the value of $r_{\text{max}}$. The left
  plot is evaluated with $\mathfrak{Z}$, so the small $r$ region includes most
  of the contribution. The right plot is evaluated with $\mathfrak{Z}'$ in
  place of $\mathfrak{Z}$, so the QCD finite volume is better controlled in
  the small $r$ region.}
\end{figure}

\vspace{-0.3cm}

We can see that by using only a larger QED lattice, the major part of the
finite-volume effects have been removed. However the disagreement between the
results shown in the 2nd and 3rd lines of Table \ref{fig:confs} implies that a
$16^3$ lattice with a spatial extent of $1.8
\mathrm{\ensuremath{\operatorname{fm}}}$ is not large enough to entirely
suppress the QCD finite volume effect. This can be seen from the right plot of
Figure \ref{fig:bin-distance-16I-24I}. In the small $r$ region, where we
control the QCD finite volume effects. the result from the 16I QCD/24 QED
calculation agrees very well with 24I. However, as $| r |$ becomes larger, the
quark loop evaluated in 16I is affected by the boundary and begins to deviate
from the 24I results. Note because we use periodic boundary conditions for the
quark propagators, the maximum spatial separation between source and sink in
any direction is $8$ for quark propagators on the 16I lattice.

\section{Physical Pion Mass $48^3$ Lattice Simulation}

\begin{figure}[H]
  \begin{center}
    \resizebox{0.45\columnwidth}{!}{\includegraphics{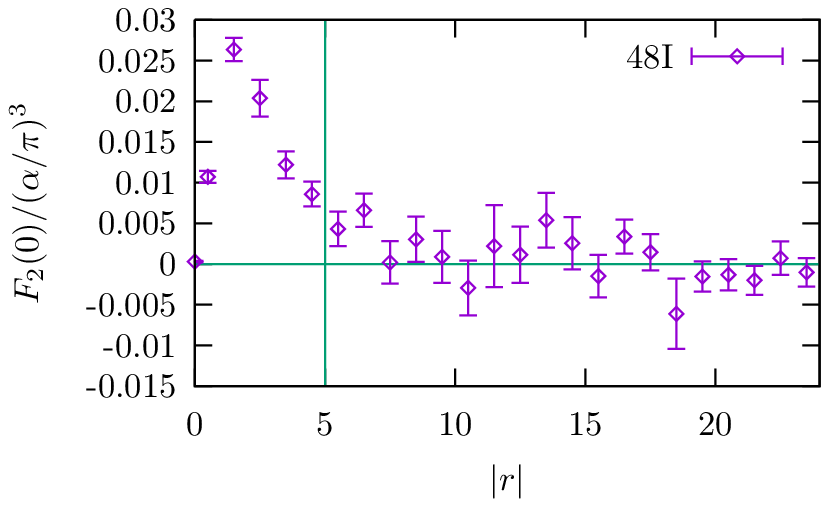}}
    \resizebox{0.45\columnwidth}{!}{\includegraphics{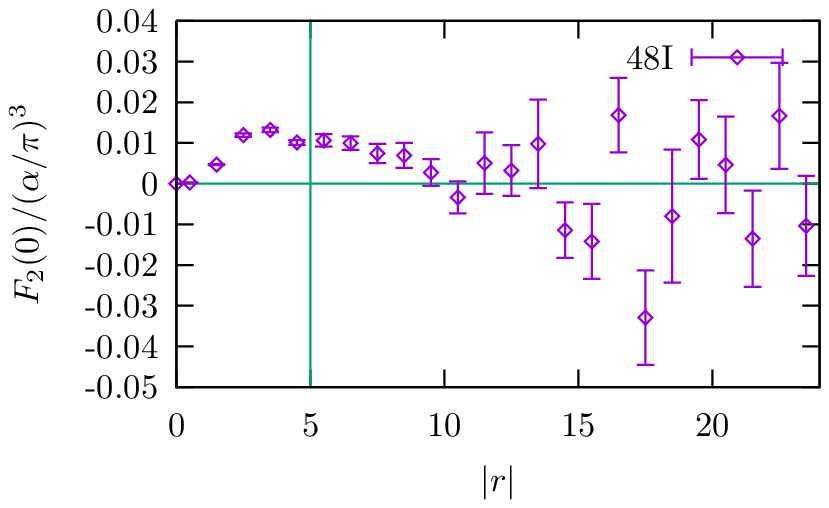}}
  \end{center}
  
\vspace{-0.8cm}
  \caption{\label{fig:bin-distance-48I}Histograms of the contribution to $F_2$
  from different separations $| r | = | x - y |$. The sum of all these points
  gives the final result for $F_2$. The vertical lines at $| r | = 5$ indicate
  the value of $r_{\text{max}}$. The left plot is evaluated with
  $\mathfrak{Z}$, so the small $r$ region includes most of the contribution.
  The right plot is evaluated with $\mathfrak{Z}'$ in place of $\mathfrak{Z}$,
  so the QCD finite volume effects are better controlled in the small $r$
  region. Currently we only have $8$ configurations to make the plot on the
  right.}
\end{figure}

Here we report on our ongoing numerical study, being done in a larger
collaboration, on a $48^3$ lattice with a physical pion mass, and a spatial
extent of $5.5 \mathrm{\ensuremath{\operatorname{fm}}}$ {\cite{Blum:2014tka}}.
The computation has so far been completed on $30$ configurations separated by
$40$ MD time units. The AMA {\cite{Blum:2012uh}} technique is used to reduce
the statistical error. The small AMA correction term is given in Table
\ref{fig:conf-48I} with the label AMA, and has already been added to the total
contribution listed in the table. We also plot the histograms in Figure
\ref{fig:bin-distance-48I}, which just contain results from the imprecise
solves.\vspace{-0.2cm}

\begin{table}[H]
  \begin{center}
    \begin{tabular}{cccccc}
      Ensemble & QCD Size & QED Size & $t_{\text{snk}} - t_{\text{src}}$ &
      $\frac{F_2 (q^2 = 0)}{ (\alpha / \pi)^3}$ & AMA\\
      48I {\cite{Blum:2014tka}} & $48^3 \times 96$ & $48^3 \times 96$ & $40$ &
      $0.0926 (124)$ & $0.0008 (23)$\\
      48I {\cite{Blum:2014tka}} & $48^3 \times 96$ & $48^3 \times 96$ & $48$ &
      $0.0946 (131)$ & $0.0007 (24)$
    \end{tabular}
  \end{center}
  
\vspace{-0.6cm}
  \caption{\label{fig:conf-48I}Finite volume effects studies. $a^{- 1} = 1.73
  \mathrm{\ensuremath{\operatorname{GeV}}}$, $m_{\pi} = 139
  \mathrm{\ensuremath{\operatorname{MeV}}}$, $m_{{\mu}} = 106
  \mathrm{\ensuremath{\operatorname{MeV}}}$.}
\end{table}

\vspace{-0.3cm}

The results for the two different time separations are quite close suggesting
that the effect of excited states is under control. Based on our finite volume
and finite lattice spacing in our pure QED simulations in Ref
{\cite{Jin:2015eua,Blum:2015gfa}}, we would estimate the above result
could have 20\% discretization errors and significant finite volume errors. As
a result, the infinite volume, continuum value for the connected
light-by-light contribution could be twice as large as this value. However,
the disconnected light-by-light diagrams may contribute negatively and cancel
part of the above enhancement.\vspace{-0.2cm}

\vspace{-0.3cm}

The plot on the right of Figure \ref{fig:bin-distance-48I} suggest that the
QCD finite volume effects will likely to be small, because it seems that the
contribution vanishes rather quickly at $| r | \sim 10$, much smaller than
$24$, which is half of the spatial size of the lattice. However, more
statistics are needed to draw a firm conclusion.

\section{Conclusions and Acknowledgments}

We have described two sampling strategies for our choice of stochastic
integration points. We have also presented a new lattice-based numerical
method for reducing the QED power-law, finite volume error by introducing a
second, larger ``QED volume'' in which the photon and muon parts of the path
integral are evaluated. This method has been illustrated by comparing HLbL
calculated on combinations of $16^3$, $24^3$ volumes. Finally we have shown
our current result using this method on a physical-pion-mass, $48^3$, $5.5
\mathrm{\ensuremath{\operatorname{fm}}}$ lattice. We plan to a) address the
discretization errors by computing on our finer, physical-pion $64^3$ lattice
with similar physical volume. b) address the finite volume effect by using the
$48^3$ QCD box inside a larger QED box and c) compute disconnected diagrams
within the framework of this newly-developed evaluation strategy.

We would like to thank our RBC and UKQCD collaborators for helpful discussions
and support. We would also like to thank RBRC for BG/Q computer time. The
$48^3$ computation is performed on Mira with ALCC allocation using BAGEL
{\cite{Boyle:2009vp}} library. T.B is supported by U.S. DOE grant
\#DE-FG02-92ER41989. N.H.C and L.C.J are supported by U.S. DOE grant
\#DE-SC0011941. M.H is supported by Grants-in-Aid for Scientific Research
\#25610053. T.I and C.L are supported by U.S. DOE Contract
\#AC-02-98CH10996(BNL). T.I is also supported by Grants-in-Aid for Scientific
Research \#26400261. This research used resources of the Argonne Leadership
Computing Facility, which is a DOE Office of Science User Facility supported
under Contract DE-AC02-06CH11357.

\bibliographystyle{hplain}
\bibliography{ref.bib}

\end{document}